\documentclass[sigconf]{acmart}

\usepackage{booktabs} 
\usepackage{graphicx}
\usepackage{placeins}
\usepackage{balance}
\usepackage{xspace}
\usepackage{paralist}

\usepackage{tikz}
\usepackage{pgfplots}
\pgfplotsset{compat=1.18}
\usetikzlibrary{plotmarks,shapes,pgfplots.dateplot}
\usetikzlibrary{calc,trees,positioning,arrows,chains,shapes.geometric,%
    decorations.pathreplacing,decorations.pathmorphing,shapes,%
    matrix,shapes.symbols}
\usetikzlibrary{3d,decorations.text,shapes.arrows,fit,backgrounds,quotes,arrows.meta}

\colorlet{color01}{red!10!white}
\colorlet{color02}{orange!10!white}
\colorlet{color03}{green!10!white}
\colorlet{color04}{blue!10!white}
\colorlet{color05}{magenta!20!white}
\colorlet{color06}{black!15!white}
\colorlet{color07}{yellow!10!white}
\colorlet{color08}{white}
\colorlet{color1}{red!80!black}
\colorlet{color1a}{red!50!white}
\colorlet{color2}{orange!80!black}
\colorlet{color2a}{orange!50!white}
\colorlet{color3}{green!50!black}
\colorlet{color3a}{green!90!black}
\colorlet{color4}{blue!80!black}
\colorlet{color4a}{blue!50!white}
\colorlet{color5}{magenta!80!black}
\colorlet{color5a}{magenta!50!white}
\colorlet{color6}{black!80!black}
\colorlet{color6a}{black!50!white}
\colorlet{color7}{white!30!black}
\colorlet{color7a}{black!50!white}
\colorlet{color8}{yellow}
\colorlet{color8a}{black!50!yellow}

\tikzset{every picture/.style={semithick},every path/.style={thick,rounded corners,->}}
\tikzset{
  ne/.style={ draw=none, fill=none, font=\footnotesize\sffamily,  minimum height=0em, text centered},
  ncirc/.style={ circle, draw=black, thin, fill=none, font=\footnotesize\sffamily,  minimum height=2em, inner sep=0, text centered},
  nd/.style={ font=\sffamily,  text centered},
  nodecomp/.style={ rectangle,  rounded corners,  draw=black, thick, text width=2em,  font=\footnotesize\sffamily, minimum height=1.3em,  text centered},
  nodevar/.style={ nodecomp, fill=green!10,
  },
  diablo/.style={ rectangle,  rounded corners,  draw=black, thick, text width=10em,  font=\footnotesize\sffamily, minimum height=1em,  text centered},
  branch/.style ={circle,inner sep=0pt,minimum size=1.5mm,fill=black,draw=black},
  diablo2/.style={ rectangle,  rounded corners,  fill=red!10, draw=black!80,thick, text width=3em,  font=\footnotesize\sffamily, minimum height=2.7em,  text centered},
  diafeature/.style={ rectangle, rounded corners=2pt,  fill=green!10, draw=black!80,thick, text width=1em,  font=\footnotesize\sffamily, minimum height=6em,  text centered},
  diafeatnarr/.style={ rectangle, rounded corners=2pt,  fill=green!10, draw=black!80,thick, text width=.5em,  font=\footnotesize\sffamily, minimum height=6em,  text centered},
  dialoss/.style={ diablo2, fill=green!10,
  },
  rotnode/.style={ anchor=center, rotate=90, font=\footnotesize\sffamily
  },
  diaext/.style={ diablo2,fill=yellow!40, },
  diablo3/.style={rectangle, rounded corners, fill=blue!10, draw=blue!40,thick, text width=3.5em,  font=\footnotesize\sffamily\bfseries, text=blue, minimum height=1.5em, text centered},
  line/.style={draw=red,rounded corners,thick, ->, decoration={markings,mark=at position 1 with {\arrow[scale=4,>=stealth]{>}}},postaction={decorate}},
  element/.style={ tape, top color=white, bottom color=blue!50!black!60!, minimum width=8em, draw=blue!40!black!90, very thick, text width=10em, minimum height=3.5em, text centered, on chain},
  every join/.style={->,rounded corners,thick,shorten >=1pt},  decoration={brace},
  lineblue/.style={    join,line width=.07cm,->,blue!20  }
}

\copyrightyear{2023} 
\acmYear{2023} 
\setcopyright{acmlicensed}\acmConference[MM '23]{Proceedings of the 31st ACM International Conference on Multimedia}{October 29-November 3, 2023}{Ottawa, ON, Canada}
\acmBooktitle{Proceedings of the 31st ACM International Conference on Multimedia (MM '23), October 29-November 3, 2023, Ottawa, ON, Canada}
\acmPrice{15.00}
\acmDOI{10.1145/3581783.3612817}
\acmISBN{979-8-4007-0108-5/23/10}

\usepackage{booktabs}       
\usepackage{amsfonts}       
\usepackage{nicefrac}       
\usepackage{microtype}      
\usepackage{lipsum}
\usepackage{graphicx}
\usepackage{paralist}
\usepackage{caption}

\usepackage{tikz}
\usepackage{pgfplots}
\usetikzlibrary{plotmarks,shapes,snakes,pgfplots.dateplot}
\usetikzlibrary{calc,trees,positioning,arrows,chains,shapes.geometric,%
    decorations.pathreplacing,decorations.pathmorphing,shapes,%
    matrix,shapes.symbols,shapes.geometric}
\usetikzlibrary{3d,decorations.text,shapes.arrows,fit,backgrounds,quotes,arrows.meta}

\pgfdeclarelayer{background}
\pgfdeclarelayer{foreground}
\pgfsetlayers{background,main,foreground}

\tikzset{every picture/.style={semithick},every path/.style={thick,rounded corners,->}}

\tikzset{
  ne/.style={
    draw=none, fill=none,
    font=\footnotesize\sffamily, 
    minimum height=0em,
    text centered},
  ncirc/.style={
    circle,
    draw=black, fill=none,
    font=\footnotesize\sffamily, 
    minimum height=2em,
    inner sep=0,
    text centered},
  nd/.style={
    font=\sffamily, 
    text centered},
  nodecomp/.style={
    rectangle, 
    rounded corners, 
    draw=black, thick,
    text width=2em, 
    font=\footnotesize\sffamily,
    minimum height=1.3em, 
    text centered},
  nodevar/.style={
    nodecomp,
    fill=green!10,
  },
  diablo/.style={
    rectangle, 
    rounded corners, 
    draw=black, thick,
    text width=10em, 
    font=\footnotesize\sffamily,
    minimum height=3em, 
    text centered},
  branch/.style ={circle,inner sep=0pt,minimum size=1.5mm,fill=black,draw=black},
  diablo2/.style={
    rectangle, 
    rounded corners, 
    fill=red!10,
    draw=black!80,thick,
    text width=13em, 
    font=\footnotesize\sffamily,
    minimum height=1.5em, 
    text centered},
  diafeature/.style={
    rectangle,
    rounded corners=2pt, 
    fill=green!10,
    draw=black!80,thick,
    text width=1em, 
    font=\footnotesize\sffamily,
    minimum height=6em, 
    text centered},
  diafeatnarr/.style={
    rectangle,
    rounded corners=2pt, 
    fill=green!10,
    draw=black!80,thick,
    text width=.5em, 
    font=\footnotesize\sffamily,
    minimum height=6em, 
    text centered},
  dialoss/.style={
    diablo2,
    fill=green!10,
  },
  rotnode/.style={
    anchor=center, rotate=90, font=\footnotesize\sffamily
  },
  diaext/.style={
    diablo2,
    fill=yellow!40,
  },
  diablo3/.style={
    rectangle, 
    rounded corners, 
    fill=blue!10,
    draw=blue!40,thick,
    text width=3.5em, 
    font=\footnotesize\sffamily\bfseries,
    text=blue,
    minimum height=1.5em,
    text centered},
  line/.style={draw=red,rounded corners,thick, ->, decoration={markings,mark=at position 1 with %
    {\arrow[scale=4,>=stealth]{>}}},postaction={decorate}},
  element/.style={
    tape,
    top color=white,
    bottom color=blue!50!black!60!,
    minimum width=8em,
    draw=blue!40!black!90, very thick,
    text width=10em, 
    minimum height=3.5em, 
    text centered, 
    on chain},
  every join/.style={->,rounded corners,thick,shorten >=1pt},
  decoration={brace},
  lineblue/.style={
  	join,line width=.07cm,->,blue!20
  }
}

\tikzset{pics/fake box/.style args={
#1 with dimensions #2 and #3 and #4}{
code={
  \draw[gray,ultra thin,rounded corners=0pt,fill=#1]  (0,0,0) coordinate(-front-bottom-left) to
  ++ (0,#3,0) coordinate(-front-top-right) --++
  (#2,0,0) coordinate(-front-top-right) --++ (0,-#3,0) 
  coordinate(-front-bottom-right) -- cycle;
  \draw[gray,ultra thin,rounded corners=0pt,fill=#1] (0,#3,0)  --++ 
   (0,0,#4) coordinate(-back-top-left) --++ (#2,0,0) 
   coordinate(-back-top-right) --++ (0,0,-#4)  -- cycle;
  \draw[gray,ultra thin,rounded corners=0pt,fill=#1!80!black] (#2,0,0) --++ (0,0,#4) coordinate(-back-bottom-right)
  --++ (0,#3,0) --++ (0,0,-#4) -- cycle;
}
}}
\tikzset{circle dotted/.style={dash pattern=on .05mm off 2mm, line cap=round}}

\author{Qi Yang}
\email{yangqi@itmo.ru}
\affiliation{%
   \institution{ITMO University}
   \city{Saint Petersburg}
   \country{Russia}
   }

\author{Marlo Ongpin}
\email{marlo@somin.ai}
\affiliation{%
   \institution{SoMin.ai Research}
   \city{Singapore}
   \country{Singapore}}

\author{Sergey Nikolenko}
\email{sergey@logic.pdmi.ras.ru}
\affiliation{%
   \institution{ITMO University}
   \institution{Steklov Institute of Mathematics}
   \city{Saint Petersburg}
   \country{Russia}
   }

\author{Alfred Huang}
\email{alfred@somin.ai}
\affiliation{%
   \institution{SoMin.ai Research}
   \city{Singapore}
   \country{Singapore}}

\author{Aleksandr Farseev}
\email{sasha@somin.ai}
\affiliation{%
   \institution{SoMin.ai Research}
   \city{Singapore}
   \country{Singapore}}

\vspace{-1cm}


\begin{document}

\title{Against Opacity: Explainable AI and Large Language Models for Effective Digital Advertising}

\begin{abstract}
The opaqueness of modern digital advertising, exemplified by platforms such as \emph{Meta Ads}, raises concerns regarding their autonomous control over audience targeting, pricing structures, and ad relevancy assessments. Locked in their leading positions by network effects, ``Metas and Googles of the world'' attract countless advertisers who rely on intuition, with billions of dollars lost on ineffective social media ads. The platforms' algorithms use huge amounts of data unavailable to advertisers, and the algorithms themselves are opaque as well. This lack of transparency hinders the advertisers' ability to make informed decisions and necessitates efforts to promote transparency, standardize industry metrics, and strengthen regulatory frameworks. In this work, we propose novel ways to assist marketers in optimizing their advertising strategies via machine learning techniques designed to analyze and evaluate content, in particular, predict the click-through rates (CTR) of novel advertising content. Another important problem is that large volumes of data available in the competitive landscape, e.g., competitors' ads, impede the ability of marketers to derive meaningful insights. This leads to a pressing need for a novel approach that would allow us to summarize and comprehend complex data. Inspired by the success of ChatGPT in bridging the gap between large language models (LLMs) and a broader non-technical audience, we propose a novel system that facilitates marketers in data interpretation, called SODA, that merges LLMs with explainable AI, enabling better human-AI collaboration with an emphasis on the domain of digital marketing and advertising. By combining LLMs and explainability features, in particular modern text-image models, we aim to improve the synergy between human marketers and AI systems.

\end{abstract}



\begin{teaserfigure}\centering
  \includegraphics[width=.9\textwidth]{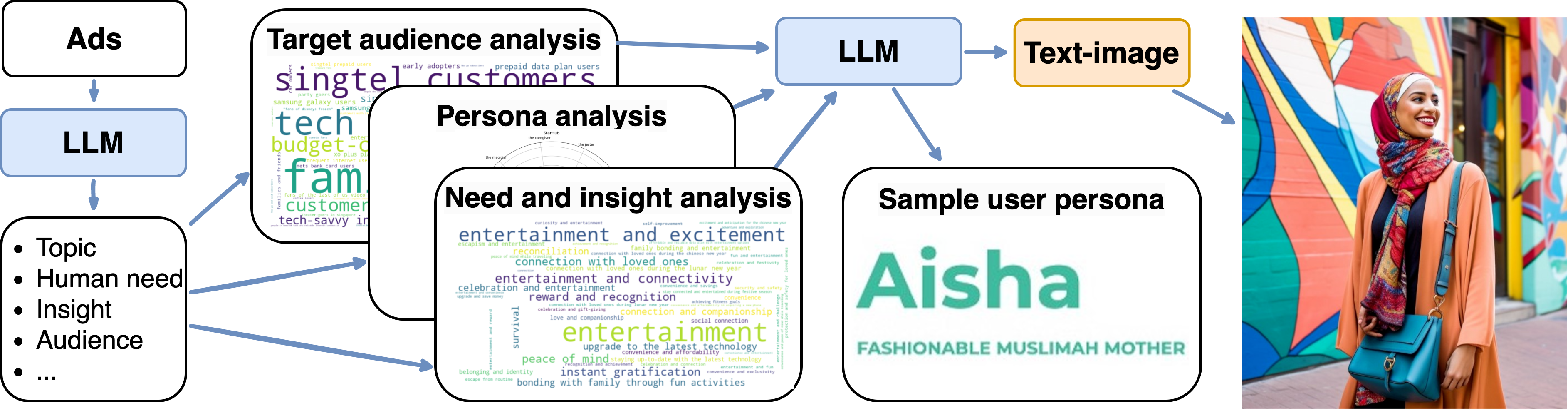}
  \caption{Overview of the LLM-based advertising analysis framework SODA (Section~\ref{sec:llm}).}
  \label{fig:teaser}
\end{teaserfigure}

\maketitle

\section{Introduction}

The online advertising industry is the poster child of data science. \emph{Google} and \emph{Facebook} became industry-dominating behemoths to a large extent because they excelled at crunching the numbers and showing the best online ads to their primary assets, user audiences, while \emph{Amazon} did the same for item recommendations in its online store. In academia, the \emph{Netflix Prize Competition}~\cite{netflix} devoted to movie recommendations was one of the first open competitions with serious prizes and organization, a pioneer that would eventually lead to \emph{Kaggle} and innumerable open leaderboards that nowadays track the state of the art in virtually every measurable ML task. The Netflix Prize itself has led to significant breakthroughs in collaborative filtering, and its dataset is still used as one of the standard benchmarks~\cite{netflix}. One definitely cannot say that the field of recommender systems, in particular online advertising, lacks the attention of machine learning researchers, and many important advances keep being made every month~\cite{10.5555/1941884, nie1,nie2,nie3,rec2,rec3,rec4,10.1145/3336191.3371831}. 

However, most advances are being made on the side of the platforms (ad marketplaces) such as \emph{Facebook} (\emph{Meta})~\cite{fb-ctr-practical-lessons}, \emph{Google}~\cite{goog-ctr-view-from-trenches}, \emph{Alibaba}~\cite{baba-rep-learning-ctr}, or \emph{Taobao}~\cite{baba-image-matters}, and therefore they are not accessible to the advertising platform users, i.e., digital marketers. Collaborative filtering datasets are understandably private, and marketing professionals that create advertising content do not have access to the data needed to predict their own future performance. Note that these predictions are often self-fulfilling: if, e.g., \emph{Meta} models predict low click-through ratio (CTR) for your ad, \emph{Meta} will charge you more for showing it, probably show it less, and the campaign will likely be a failure regardless of how accurate the CTR prediction has been~\cite{fb_ctr}. Often, there is no practical way to control the cost of advertising; technically, if a platform decided to charge more money for an ad nothing could prevent them from doing so. 

Moreover, even if marketing professionals could run the corresponding models, that would only be of modest help with their job, which is \emph{content creation}. Suppose that a model tells you that your new ad is a bad match for your audience, and the expected CTR is low. How do you fix that? It cannot be a pure collaborative filtering model since it has to predict CTR for a new ad that has not been shown to users yet, but it is still an opaque model that maps your ad content into a latent representation via ``giant inscrutable matrices''. So all you can do even if you have such a model is to try and make a different ad, get a new prediction, and work via trial and error.

One potential way to address this issue involves visualizing the decision-making process of a neural network, providing marketers with insights into the rationale behind specific predictions made by AI models~\cite{10.1145/3503161.3548769,we1,we2,niko1}. Therefore, our first contribution in this work is a new variation of a state-of-the-art CTR prediction model coupled with a mechanism for analyzing the ad images (banners) via an image attention mechanism. The results provide human-understandable analysis that can be turned into actionable insights. 

However, this is only the beginning. 
Individual ad analysis via explainable ML models has proven beneficial in scenarios such as individual content evaluation prior to starting an advertising campaign, but it is much less practical when applied to large volumes of images and text ads in real-world settings. The time constraints faced by marketers impede their ability to effectively process and extract key content traits in their own advertising practices.

In our opinion, the long-awaited revolution in digital advertising and content marketing will occur when both the ads themselves and the results of opaque models can be explained in ways that are both understandable for humans and actionable in terms of business results. We believe that the time for this revolution is now, and in this work we show that large language models such as GPT-3.5~\cite{ouyang2022training} and GPT-4~\cite{openai2023gpt4} are already increasingly able to explain the ``reasoning'' behind recommender models and provide aggregate insights about advertising campaigns consisting of hundreds of individual ads. Prior to LLMs, approaches to aggregate text corpora in the context of recommender systems had been proposed via topic modeling~\cite{10.1145/2615569.2615680,10.1007/978-3-319-27101-9_5}, sometimes coupled with deep learning~\cite{TN18} and user profiling~\cite{farseev2,TN17}, but topic modeling is based on the bag-of-words assumption and cannot summarize text as an LLM does; visual understanding of ads had also been explored with convolutional networks~\cite{savchenko-etal-2020-ad}.

Therefore, our main contribution is that we present preliminary results for a road-map that could achieve this holy grail of content marketing: provide \emph{explainable}, \emph{actionable} insights into advertising content along with possible strategies for improvement with models that could work on the side of a small advertising agency rather than a huge platform. We begin with direct CTR prediction and then proceed to provide explainable insights and content recommendations with large language models and even visual generative AI (see Fig.~\ref{fig:teaser}).

The paper is organized as follows: in Section~\ref{sec:ctr}, we present an improved model for CTR prediction and visualization procedures for advertising banners, Section~\ref{sec:llm} introduces our approach to explainable ad analysis with large language models, Section~\ref{sec:case} shows the results of a case study that confirms the effectiveness of our approach, and Section~\ref{sec:concl} concludes the paper.

\section{Explaining Opaque AI with AI: CTR Prediction and Visualizations}\label{sec:ctr}

The lack of transparency within the advertising sector has been widely acknowledged as a primary reason for the inefficient allocation of advertising budgets. Notably, the responsibility for determining the cost per 1,000 impressions (CPM) and selecting competing entities in a programmatic auction rests primarily with the platform (we will use \emph{Meta} as the running example). This decision-making process is in fact a result of numerous intricately interwoven machine learning (ML) models designed to dynamically match content with precise targeting criteria and individualized user profiles on \emph{Meta}. These models are instrumental in estimating the likelihood of a user engaging in specific actions within the \emph{Meta} ecosystem.

\def\sowide{\emph{SoWide-v2}\xspace}

\begin{figure*}[!t]
\begin{tabular}{cc}
\begin{minipage}{.38\linewidth}\centering
\includegraphics[width=.8\linewidth]{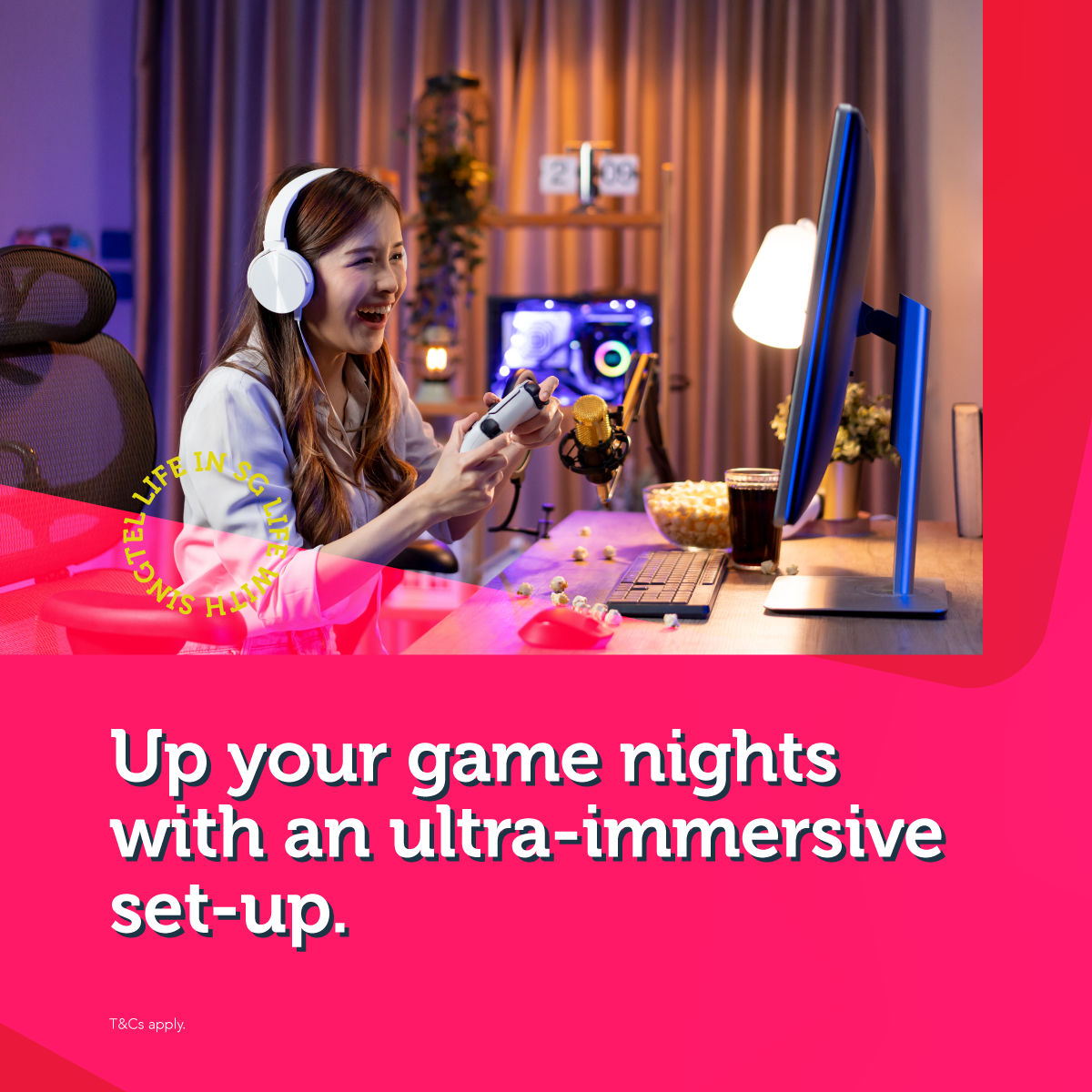} 
\vspace{-1em}
\caption{A sample advertising banner on \emph{Meta}.}\label{fig:singtel-banner}
\end{minipage}
& 
\begin{minipage}{.6\linewidth}\centering
\resizebox{.9\linewidth}{!}{
\begin{tikzpicture}[node distance=.2cm]
\def\xx{1.5}
\def\xxx{-.0}
\def\yy{.75}

\node[diablo2,minimum height=2pt,text width=4em,fill=color04] (in1) at (0, 5*\yy) {{Continuous features}};
\node[diablo2,minimum height=2pt,text width=4em,fill=color04] (in2) at (1*\xx, 5*\yy) {Categorical features};

\node[diablo2,minimum height=2pt,text width=4em,fill=color04] (in3) at (2.5*\xx, 5*\yy) {Text (creative)};
\node[diablo2,minimum height=2pt,text width=4em,fill=color04] (in4) at (4.5*\xx, 5*\yy) {Images (creative)};

\node[diablo2,minimum height=2pt,text width=8em,fill=color04] (bp) at (.5*\xx, 4*\yy) {Tabular preprocessor};
\node[diablo2,minimum height=2pt,text width=8em,fill=color04] (tp) at (2.5*\xx, 4*\yy) {Tokenizer};
\node[diablo2,minimum height=2pt,text width=8em,fill=color04] (ip) at (4.5*\xx, 4*\yy) {Image preprocessor};

\node[diablo2,minimum height=2pt,text width=8em,fill=color05] (bm) at (.5*\xx, 3*\yy) {TabTransformer};
\node[diablo2,minimum height=2pt,text width=8em,fill=color05] (tm) at (2.5*\xx, 3*\yy) {BERT};
\node[diablo2,minimum height=2pt,text width=8em,fill=color05] (im) at (4.5*\xx, 3*\yy) {ViT};

\node[diablo2,minimum height=2pt,text width=6em,fill=color01] (fc1) at (.5*\xx, 2*\yy) {FC layer};
\node[diablo2,minimum height=2pt,text width=6em,fill=color01] (fc2) at (2.5*\xx, 2*\yy) {FC layer};
\node[diablo2,minimum height=2pt,text width=6em,fill=color01] (fc3) at (4.5*\xx, 2*\yy) {FC layer};

\node[diablo2,minimum height=2pt,text width=20em,fill=color03] (concat) at (2.5*\xx, 1*\yy) {Concatenation};
\node[diablo2,minimum height=2pt,text width=10em,fill=color01] (fc) at (2.5*\xx, 0*\yy) {FC layer};
\node[diablo2,minimum height=2pt,text width=10em,fill=color03] (soft) at (2.5*\xx, -1*\yy) {Softmax};

\draw (in1.south) -- (in1.south |- bp.north);
\draw (in2.south) -- (in2.south |- bp.north);
\draw (in3) -- (tp);
\draw (in4) -- (ip);
\draw (bp) -- (bm);
\draw (tp) -- (tm);
\draw (ip) -- (im);
\draw (bm) -- (fc1);
\draw (tm) -- (fc2);
\draw (im) -- (fc3);
\draw (fc1.south) -- (fc1.south |- concat.north);
\draw (fc2.south) -- (fc2.south |- concat.north);
\draw (fc3.south) -- (fc3.south |- concat.north);
\draw (concat) -- (fc);
\draw (fc) -- (soft);
\end{tikzpicture}  
} 
\vspace{-1em}
\caption{\sowide architecture}
\label{fig:sowide}
\end{minipage}
\end{tabular}
\vspace{-1em}
\end{figure*}

As an illustration, consider a hypothetical \emph{Meta} user named Simon who is anticipated to click on an ad (perform the ``Click'' action) with the slogan ``Up your game nights with an ultra-immersive set-up'' displayed on a \emph{Meta Ad} banner showcasing \emph{Singtel}, a mobile operator company, and their home internet broadband product (Fig.~\ref{fig:singtel-banner}). This prediction is done by \emph{Meta}'s internal ML models, and quite often contradicts \emph{Meta}'s widely publicized ''best practices'' blueprints~\cite{metaBlueprints}. Here, it is crucial to acknowledge the additional information that advertising engines such as \emph{Meta} take into account. They are free to use factors such as Simon's past visits to telecom websites, pictures showing computer games in Simon's account on \emph{Meta}, and much more. Moreover, these factors include \emph{Meta}'s own revenue considerations, prediction of the ad's ``relevance'' by \emph{Meta} itself, timing of displaying this ad during the day, recency of the ad account (to incentivize new advertisers with improved performance), and the internal ``ranking'' of advertisers based on their history of disapproved ads, a process overseen by \emph{Meta}. Regrettably, these predictive estimations are further influenced by the accuracy of \emph{Meta}'s ML models that profile Simon's content. For instance, when Simon is observed putting a diaper on his child, \emph{Meta}'s object recognition system might mistakenly associate it with an ``Inflatable Boat / Fishing'' interest; this is a real-life incident on the \emph{Meta} platform, and such mistakes compound into suboptimal ad-related predictions down the line.

Confronted with numerous intricate technical hurdles, digital marketers, who frequently lack technical expertise, often resort to intuitive judgment or a trial-and-error methodology in formulating and examining their creative assets within digital advertising platforms. Thus, it becomes especially important to have comprehensive data-driven guidance, not only for optimizing outcomes but also for developing cost-effective practices.

One classical approach to providing this kind of guidance is to train an ML framework to predict the prospective performance of an advertising banner before allocating actual advertising budgets. In this section, we focus on the prediction of the click-through rate (CTR) metric, known to be closely associated with ad performance, particularly in the context of awareness and traffic advertising objectives. 
We used the recently presented \emph{SoWide} model~\cite{we2} as a sample state-of-the-art CTR prediction approach; its architecture is shown in Fig.~\ref{fig:sowide}. We updated the architecture slightly by replacing the ABN model for image processing with a Vision Transformer (ViT)~\cite{VIT}, resulting in performance improvements, so we call it \sowide.

Unlike conventional supervised learning, where a data point $(\mathbf{x}, y)$ consists of both feature vector $\mathbf{x}$ and target variable $y$, the \sowide approach incorporates data from the campaign, ad set, and potentially multiple creatives to construct the features for each ad. Data points in the model leverage text and images from all creatives together with their respective estimated performances; in case of videos, we extract keyframes to obtain multiple distinct images included as additional training data. Furthermore, we extract low-level features from tabular, textual, and visual content, resulting in a comprehensive dataset that can be used to train a model capable of predicting content performance based on information from multiple modalities. After preprocessing, extracted features serve as inputs for the click-through rate (CTR) prediction model. \sowide makes the assumption that the performance of advertisements converges to an underlying global distribution~\cite{msft-ad-pred,farseev1,farseev3}, so we normalize CTR values into categorical representations. Predicted scores indicate whether the content can be classified as ``below average'', ``average'', or ``above average'' in terms of quality. 

\begin{figure*}[!t]\centering
\setlength{\tabcolsep}{7pt}
\begin{tabular}{cc}
\begin{minipage}{.18\linewidth}\centering
\setlength{\tabcolsep}{2pt}
\def\alittlespacehere{3pt}
\captionof{table}{F1-score evaluation for CTR prediction models.}
    \label{table:evaluation}
\begin{tabular}[t]{lcccccccc}
\toprule
& All & Con-\\
& & ver- \\
& & sion \\\midrule
$k$-nearest & & \\
neighbors & 0.338 &  0.254 \\[\alittlespacehere]
Random & & \\
forest & 0.302 & 0.293 \\[\alittlespacehere]
Gradient & & \\
boosting & 0.349 & 0.262 \\[\alittlespacehere]
AdaBoost & 0.289 & 0.277 \\[\alittlespacehere]
Multilayer & & \\
perceptron & 0.654 & 0.642 \\[\alittlespacehere]
\emph{SoWide} & 0.702 & 0.660 \\[\alittlespacehere]
\sowide & \textbf{0.780} & \textbf{0.671} \\
\bottomrule
\end{tabular}    
\end{minipage}
& 
\begin{minipage}{.78\linewidth}\centering
\includegraphics[width=\linewidth]{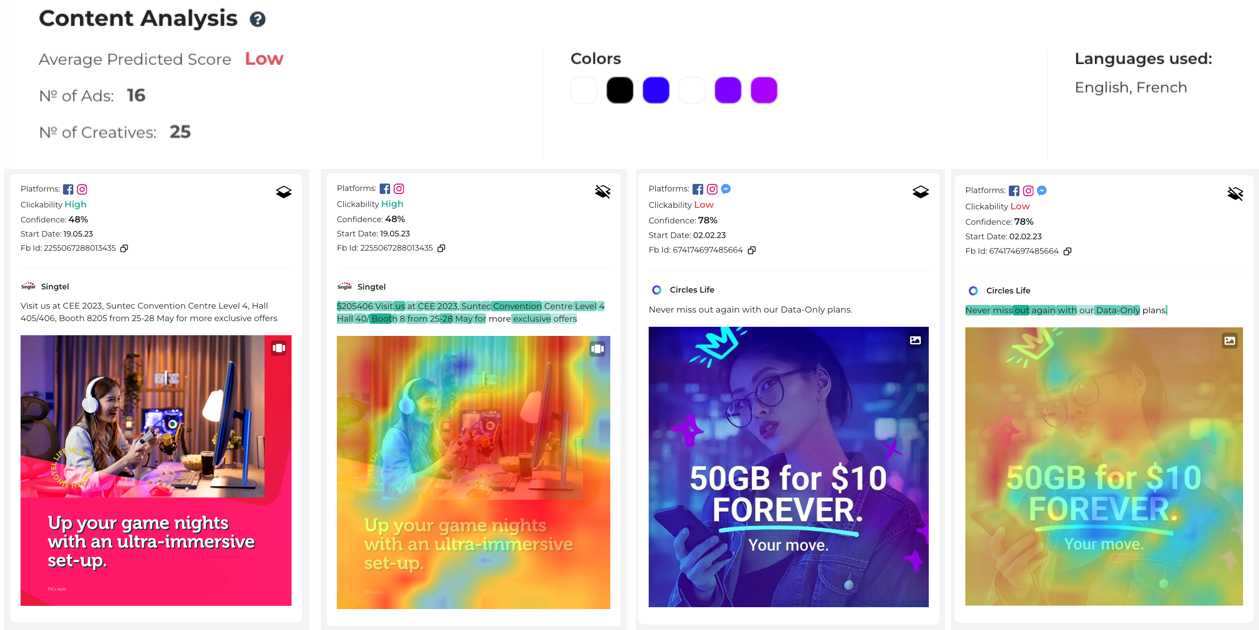}
\vspace{-2em}
\caption{Sample predicted low-CTR and high-CTR advertising banners and heatmap visualizations of the attention layers involved in the prediction.}\label{fig:visualizations}
\end{minipage}
\end{tabular}
\vspace{-1em}
\end{figure*}

In essence, \sowide is a neural network based on the ``wide and deep models'' approach well known in recommender systems~\cite{wide-and-deep}.
To facilitate representation learning for multimodal content, \sowide employs separate embedding layers and fully connected layers for each set of features. This process allows it to project sparse, high-dimensional, and low-level features into higher-level representations. To handle each modality appropriately, \sowide employs distinct deep models for feature processing. Specifically, it uses the TabTransformer~\cite{huang2020tabtransformer} for tabular features and multilingual BERT~\cite{devlin-etal-2019-bert} for textual content; the original \emph{SoWide} used the attention branch network~\cite{fukui2018cvpr} for images but for \sowide we replaced it with a Vision Transformer~\cite{VIT}. Additionally, a fully connected layer is utilized to project the sparse high-dimensional features into a denser low-dimensional representation. These representations are subsequently concatenated and fed into another fully connected layer, followed by a softmax function for CTR classification, facilitating end-to-end joint learning. The model is trained using stochastic gradient descent (SGD) for 100 epochs, and hyper-parameter optimization is performed with the tree-structured Parzen estimator~\cite{hyperopt}.

For evaluation results, we use the same datasets and baselines as the original \emph{SoWide} paper~\cite{we2}, 
comparing the performance of \sowide against the original \emph{SoWide} and several conventional machine learning baselines (there appears to be no previous work on CTR prediction before~\cite{we2} that could be used for a direct comparison) using the F$1$-score, a widely used classification metric.
Evaluation is done in two different settings: for general ad campaigns and also specifically for campaigns targeting the ``Conversion'' objective, which represents the two most prevalent and significant ad campaign objectives. The results shown in Table~\ref{table:evaluation} demonstrate that the \sowide model presents an improvement over the original \emph{SoWide}, and both models significantly outperform all classical ML baselines. Notably, the F$1$-score for the general ad campaigns reaches $0.78$, which confirms that the \sowide approach effectively accommodates the hierarchical structure inherent in advertising data, enabling effective multimodal learning for the prediction of ad performance. The results validate that \sowide is a state-of-the-art CTR prediction model.

Thus far, we have introduced a framework that enables advertisers to assess the potential performance of their own content, and potentially that of their competitors, prior to its launch. This represents a valuable tactical capability that had been unavailable to the community for a long time. However, once a creative marketer gains access to the initial prediction results for a specific content piece, another significant challenge lies in comprehending the underlying factors that contribute to its success or failure. What went wrong, what was done right, and how do we amplify the right parts while suppressing the wrong parts?

One approach to address this question would be to utilize various visualization techniques, specifically those that illustrate the decision-making process of the neural network while making a specific prediction. If the prediction is accurate, such visualizations are believed to provide insights into the underlying reasons behind the performance of a creative asset. Consequently, these visualizations can serve as a valuable resource for marketers in making informed decisions regarding the inclusion of specific components in future creative assets, enabling them to effectively communicate their requirements to the creative team.

Figure~\ref{fig:visualizations} shows an illustrative example of such visualizations. The attention layers of the neural network used for CTR prediction are visualized as interactive heatmaps, revealing the specific regions of the banner that significantly influence the model's predictions. The figure shows that such attention visualization highlights the key elements within a \emph{Singtel} banner (on the left) that contribute to its high predicted performance, namely gaming-related objects such as the monitor and the game controller. These elements effectively convey the message that a superior internet connection is essential for enhancing the gaming experience. Similarly, for the \emph{Circles.Life} banner (on the right), the areas featuring the lady in the background were found to negatively impact its performance. This suggests that the composition and balance of the banner's visual elements, particularly in relation to the overall content creation practices, may have influenced its predicted low CTR values.

\begin{figure*}[!t]\setlength{\tabcolsep}{10pt}
\begin{tabular}{cc}
\begin{minipage}{.27\linewidth}\centering
\includegraphics[width=\linewidth]{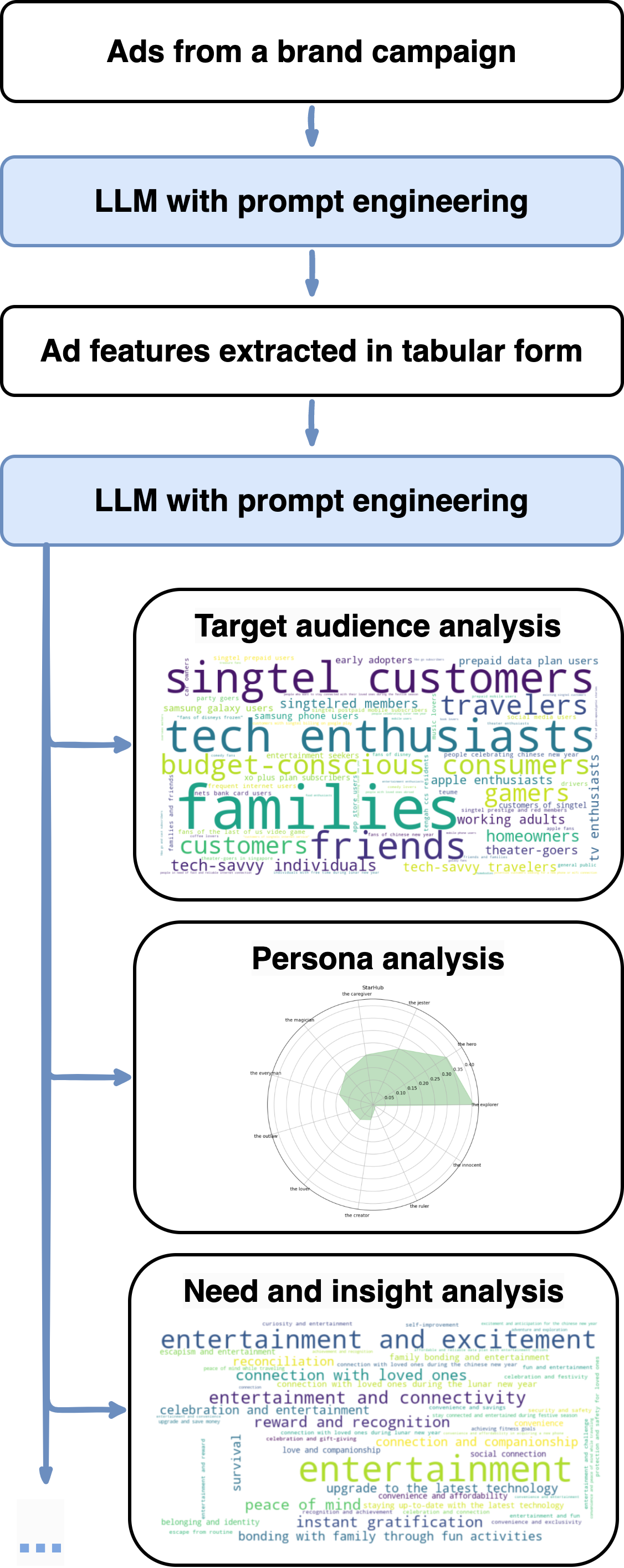} 
\caption{General pipeline of our LLM-based analysis}\label{fig:pipeline}
\end{minipage}
& 
\begin{minipage}{.68\linewidth}\centering
\includegraphics[width=\linewidth]{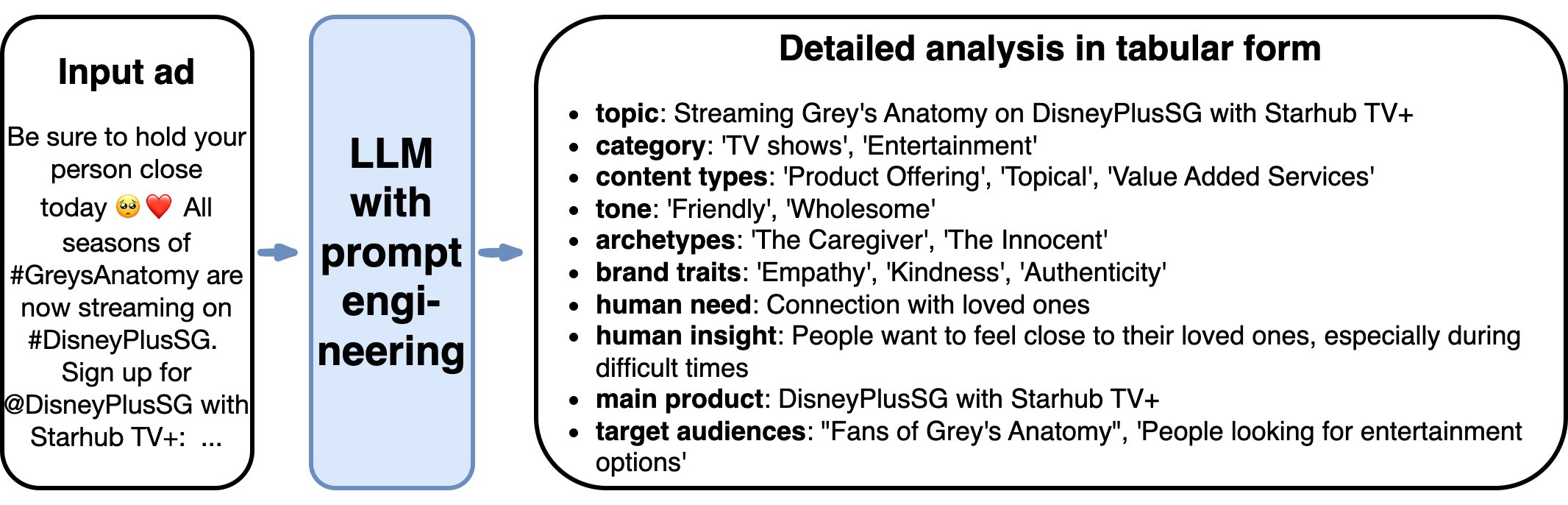}\vspace{-.3cm}

\caption{Sample ad analysis}\label{fig:adanalysis}\vspace{.2cm}

\includegraphics[width=\linewidth]{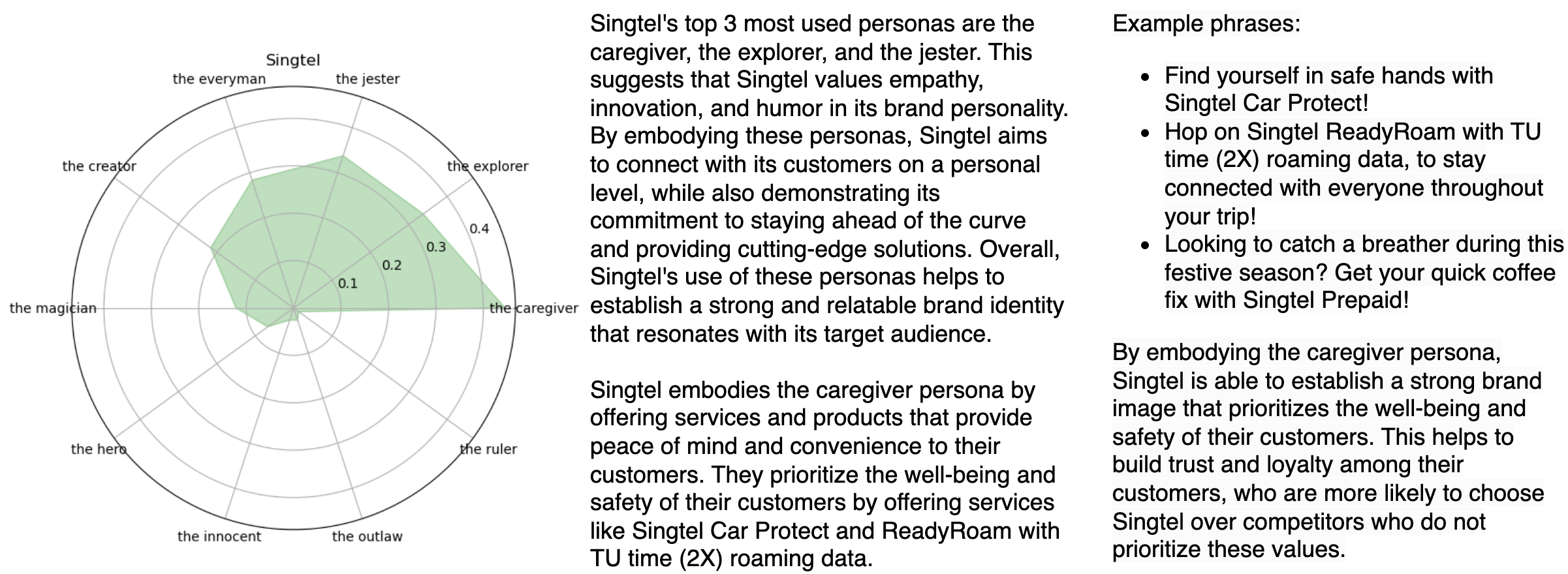}\vspace{-.3cm}

\caption{Sample brand persona analysis results}\label{fig:persona}\vspace{.2cm}

\includegraphics[width=\linewidth]{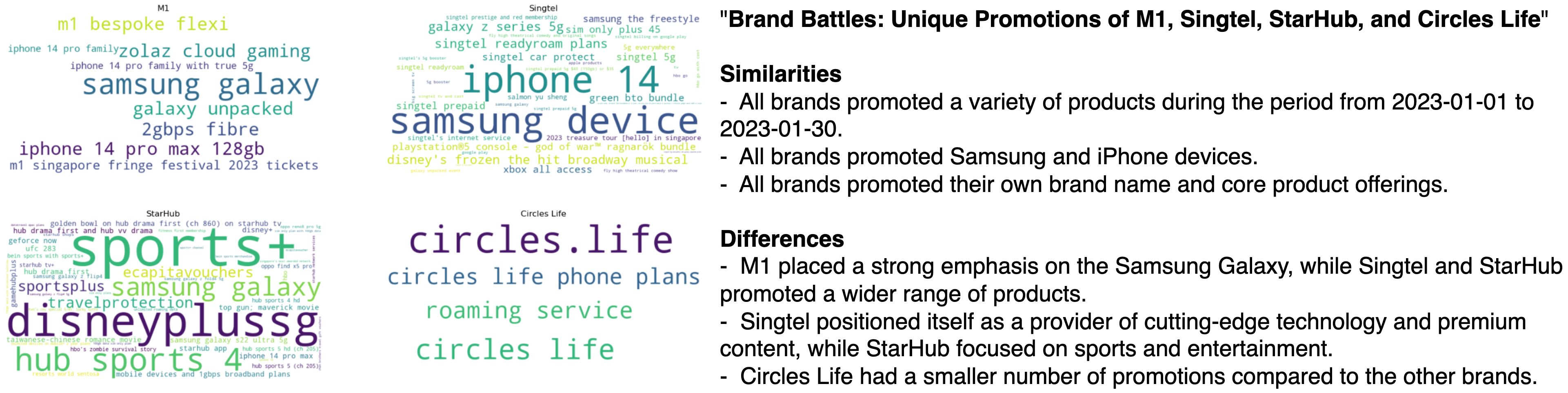}\vspace{-.3cm}

\caption{Sample brand comparative analysis results}
\label{fig:comparison}
\end{minipage}
\end{tabular}
\vspace{-1em}
\end{figure*}

\begin{figure*}[!t]
\includegraphics[width=\linewidth]{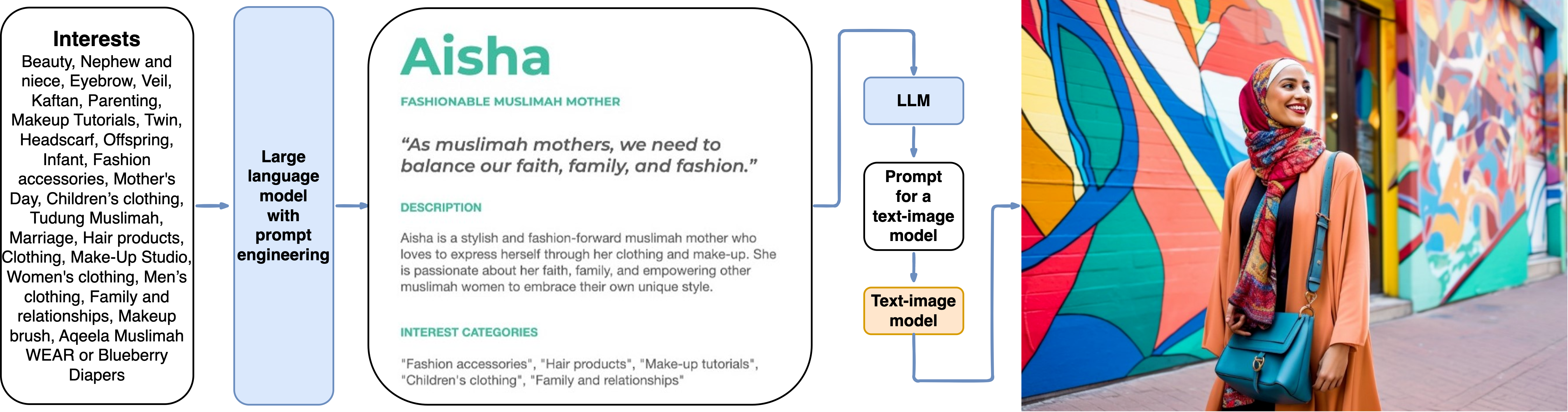}
\vspace{-2em}
\caption{Sample user persona generation results.}
\label{fig:user}
\vspace{-1em}
\end{figure*}

\section{Explaining Humans for Humans: SODA, A LLM-Based Advertising Analysis Framework}\label{sec:llm}

In the last section, we presented a system capable of effectively capturing and visualizing the factors influencing the predicted performance of ads in terms of potential CTR. However, in domains such as performance marketing decisions for choosing specific creatives for campaigns often need to be made under tight deadlines, sometimes literally in a few hours or even minutes. Moreover, these industries are characterized by large volumes of creative assets and a multitude of promotions simultaneously conducted by competitors in an ``always-on'' manner. Therefore, one cannot run detailed analysis for every ad, and there is a dire need for further automated analytical tools that would enable human marketers to rapidly comprehend available data and information.

In order to address this challenge, we present an extension to our framework with a novel approach that leverages large language models (LLMs) to provide additional insights into the data and CTR predictions, called \textit{\textbf{SODA}}. We outline an analytical pipeline that incorporates LLM-based explanations and generations and demonstrate its practical applications through a real-world scenario involving four Singapore telecommunication companies. This part of our framework aims to enhance the interpretability and comprehension of the data, facilitating better-informed decision-making in these fast-paced and competitive industries.

The general pipeline of our analysis is shown in Figure~\ref{fig:pipeline}. First, we use an LLM to extract specific well-defined insights from input ads, such as the needs served by this ad, products being advertised, and more (see below); the insights can be stored as features in tabular form. Then, we use these features together with further engineered prompts to perform generalizing analysis of a brand's target audiences, personas, needs, and insights expressed by the ads, tone, and topical categories of the current campaign and others. The resulting coverage of the campaign closely reflects campaign analysis commonly performed by marketing professionals and can be further used to tune the brand's message, tone, target audiences, personas, and more. The pipeline is also able to present specific examples helpful for marketing professionals, such as sample (imagined) user profiles or user personas, which are also one of the common marketing tools. Let us dive into some details.

Figure~\ref{fig:adanalysis} shows sample results of our initial experiments on ad analysis. We selected batches of ads from the \emph{Facebook Ad Library} for the same brand and processed them with an LLM, customized only with natural language prompt engineering. As a result, the LLM has been able to successfully identify key features of each advertisement, including excellent responses to such seemingly ``human'' questions as identifying the human need, human insight, and the main archetypes used in an ad. Moreover, answers to most questions are standardized (as the LLM was instructed) and can be subject to automated processing. This kind of analysis has always been a key part of online marketing, and to the best of our knowledge, it has never been successfully automated and scaled up before. Such tasks had always required human labeling and thus had been restricted to a few sample ads rather than the entire dataset.

As the next step, we use the ads and extracted features as inputs for a number of prompts asking to summarize information in a variety of formats commonly used in content marketing. We have seen successful summarization across the board, with important insights identified by the LLM and presented in an accessible and actionable format. Fig.~\ref{fig:persona} shows a sample result of our brand persona analysis, complete with main brand values used in the ad campaigns, the goals of using them, and detailed analysis of the primary ``caregiver'' persona, including supporting examples from the data.

Figure~\ref{fig:comparison} shows the results of a comparative analysis of four advertising campaigns run over the same time period by different brands. Again, the LLM has correctly identified its key distinguishing factors, and the list of differences is very similar to one that could be produced by a human marketing professional.

Another avenue for using state-of-the-art generative AI capabilities that we have explored is user persona generation, an important tool in content marketing that has long proven to be useful for creative work\cite{10.1145/3539597.3573031,somin_influencer,somin_responsibility}. To produce user personas, we begin with a list of interests (either extracted as shown in Fig.~\ref{fig:user} or obtained from the client and/or social media platform) and prompt the LLM to give examples of user descriptions that could fit such interests. Fig.~\ref{fig:persona} shows a sample resulting user persona, which is fully believable to the professionals. To make the result even more tangible, we supplement such user personas with images generated by a state-of-the-art text-image model, in this case, \emph{Stable Diffusion}~\cite{stablediffusion}. To make the entire pipeline self-contained we ask the original LLM to also generate the prompt for the text-image model from the user persona description and a few examples of good prompts. The results also illustrated in Fig.~\ref{fig:persona}, are very promising.

The LLM used in all experiments was ChatGPT based on GPT-3.5~\cite{ouyang2022training}, and we believe that simply switching to more powerful LLMs such as GPT-4~\cite{openai2023gpt4} may lead to further increased performance across all applications. Note also that while GPT-3.5 can only process text ads, GPT-4 is already able to analyze images jointly with text (this ability has not yet been made public at the time of writing), which is arguably even more important for content marketing.

\vspace{-1em}

\section{Case study}\label{sec:case}

To evaluate the practical value and viability of the proposed framework expansion using large language models (LLMs) for generating rapid insights and enabling prompt marketing-related decision-making, we have engaged 12 marketing professionals currently employed at marketing departments of Business-to-Consumer (B2C) brands or advertising and marketing agencies across Singapore, China, and the UK. These professionals were selected based on their extensive experience, averaging 9 years, in managing digital marketing campaigns across various industries.

The professionals were presented with the details and results of our preliminary experiments analyzing and comparing the marketing campaigns of four major telecommunication companies in Singapore, as described in Section~\ref{sec:llm}. They were then asked to provide their perspectives on the usefulness, quality, and potential impact of the insights and outputs generated by our framework.

All 12 professionals responded very positively about the value of our approach. They found high-level overviews of brand positioning and audience targeting strategies, enriched with specific examples, to be highly useful to gain quick familiarity with brand messaging and inspire new creative directions. The generated user personas and accompanying AI-generated images were praised for bringing additional richness and tangibility to the insights.

Several professionals commented that the coherent, standardized format of the outputs would allow for efficient processing and decision-making, especially given the tight timeframes frequently faced in the industry. More senior professionals have expressed that they foresee solutions like ours significantly augmenting and accelerating essential marketing functions through the automation of repetitive, labor-intensive tasks.

This highly encouraging feedback from advertising and marketing professionals suggests strong potential business value in developing and applying AI-powered solutions, such as the proposed extension of our framework, for the automation of marketing campaign analysis and strategic planning. While adoption may face initial resistance, especially from very senior professionals, many in the industry seem poised to welcome AI augmenting and enhancing their work. Our approach, focused on mimicking established human processes and outputs, appears well-suited to addressing common pain points and unlocking new efficiencies, especially in such fast-paced domains as performance marketing. 


\section{Conclusion}\label{sec:concl}

In this work, we have presented a novel advertising analysis framework, called \textit{\textbf{SODA}}, which amalgamates large language models, explainable artificial intelligence, and attention map visualization techniques, heralding a potential future of human-AI collaboration within the realm of digital advertising. Through the integration of LLMs and the incorporation of explainability aspects, our novel approach envisions enhanced efficiency and synergy between marketers and AI systems, hopefully leading to a new era of intelligent decision-making. We believe that our approach holds the promise of empowering a new generation of marketers to leverage advanced AI technologies effectively, fostering a deeper understanding of the underlying mechanisms driving ad performance and facilitating informed decision-making processes. Note that while we already show promising results, these are mostly preliminary experiments, and we strongly believe that this direction of research will bring many new advances in the nearest future.
\vspace{1em}

\section{Acknowledgement}
This work was funded by the Russian Science Foundation grant № 22-11-00135 https://rscf.ru/en/project/22-11-00135/

\bibliographystyle{ACM-Reference-Format}
\bibliography{sample-base}


\begin{thebibliography}{40}


\ifx \showCODEN    \undefined \def \showCODEN     #1{\unskip}     \fi
\ifx \showDOI      \undefined \def \showDOI       #1{#1}\fi
\ifx \showISBNx    \undefined \def \showISBNx     #1{\unskip}     \fi
\ifx \showISBNxiii \undefined \def \showISBNxiii  #1{\unskip}     \fi
\ifx \showISSN     \undefined \def \showISSN      #1{\unskip}     \fi
\ifx \showLCCN     \undefined \def \showLCCN      #1{\unskip}     \fi
\ifx \shownote     \undefined \def \shownote      #1{#1}          \fi
\ifx \showarticletitle \undefined \def \showarticletitle #1{#1}   \fi
\ifx \showURL      \undefined \def \showURL       {\relax}        \fi
\providecommand\bibfield[2]{#2}
\providecommand\bibinfo[2]{#2}
\providecommand\natexlab[1]{#1}
\providecommand\showeprint[2][]{arXiv:#2}

\bibitem[fb_({[n.\,d.]})]%
        {fb_ctr}
 \bibinfo{year}{[n.\,d.]}\natexlab{}.
\newblock \bibinfo{title}{Best practices to potentially reduce cost per result
  for Meta ads}.
\newblock
  \bibinfo{howpublished}{\url{https://www.facebook.com/business/help/321695409726523}}.
\newblock


\bibitem[net({[n.\,d.]})]%
        {netflix}
 \bibinfo{year}{[n.\,d.]}\natexlab{}.
\newblock \bibinfo{title}{Netflix Prize data}.
\newblock
  \bibinfo{howpublished}{\url{https://www.kaggle.com/datasets/netflix-inc/netflix-prize-data}}.
\newblock


\bibitem[Alekseev et~al\mbox{.}(2022)]%
        {niko1}
\bibfield{author}{\bibinfo{person}{Anton Alekseev}, \bibinfo{person}{Elena
  Tutubalina}, \bibinfo{person}{Sejeong Kwon}, {and} \bibinfo{person}{Sergey
  Nikolenko}.} \bibinfo{year}{2022}\natexlab{}.
\newblock \showarticletitle{Near-Zero-Shot Suggestion Mining with a Little Help
  from WordNet}. In \bibinfo{booktitle}{\emph{Analysis of Images, Social
  Networks and Texts}}. \bibinfo{publisher}{Springer International Publishing},
  \bibinfo{address}{Cham}, \bibinfo{pages}{23--36}.
\newblock


\bibitem[Bergstra et~al\mbox{.}(2011)]%
        {hyperopt}
\bibfield{author}{\bibinfo{person}{J. Bergstra}, \bibinfo{person}{R. Bardenet},
  \bibinfo{person}{Y. Bengio}, {and} \bibinfo{person}{B. K\'{e}gl}.}
  \bibinfo{year}{2011}\natexlab{}.
\newblock \showarticletitle{Algorithms for Hyper-Parameter Optimization}. In
  \bibinfo{booktitle}{\emph{Advances in Neural Information Processing
  Systems}}, Vol.~\bibinfo{volume}{24}. \bibinfo{publisher}{Curran Associates,
  Inc.}
\newblock


\bibitem[Buraya et~al\mbox{.}(2018)]%
        {farseev2}
\bibfield{author}{\bibinfo{person}{K Buraya}, \bibinfo{person}{A Farseev},
  {and} \bibinfo{person}{A Filchenkov}.} \bibinfo{year}{2018}\natexlab{}.
\newblock \showarticletitle{Multi-view personality profiling based on
  longitudinal data}.
\newblock \bibinfo{journal}{\emph{Lecture Notes in Computer Science}}
  \bibinfo{volume}{11018} (\bibinfo{year}{2018}), \bibinfo{pages}{15--27}.
\newblock


\bibitem[Chen et~al\mbox{.}(2023)]%
        {rec2}
\bibfield{author}{\bibinfo{person}{Jiawei Chen}, \bibinfo{person}{Hande Dong},
  \bibinfo{person}{Xiang Wang}, \bibinfo{person}{Fuli Feng},
  \bibinfo{person}{Meng Wang}, {and} \bibinfo{person}{Xiangnan He}.}
  \bibinfo{year}{2023}\natexlab{}.
\newblock \showarticletitle{Bias and Debias in Recommender System: A Survey and
  Future Directions}.
\newblock \bibinfo{journal}{\emph{ACM Trans. Inf. Syst.}} \bibinfo{volume}{41},
  \bibinfo{number}{3}, Article \bibinfo{articleno}{67} (\bibinfo{date}{feb}
  \bibinfo{year}{2023}), \bibinfo{numpages}{39}~pages.
\newblock
\showISSN{1046-8188}


\bibitem[Cheng et~al\mbox{.}(2016)]%
        {wide-and-deep}
\bibfield{author}{\bibinfo{person}{H.-T. Cheng}, \bibinfo{person}{L. Koc},
  \bibinfo{person}{J. Harmsen}, \bibinfo{person}{T. Shaked},
  \bibinfo{person}{T. Chandra}, \bibinfo{person}{H. Aradhye},
  \bibinfo{person}{G. Anderson}, \bibinfo{person}{G. Corrado},
  \bibinfo{person}{W. Chai}, \bibinfo{person}{M. Ispir}, \bibinfo{person}{R.
  Anil}, \bibinfo{person}{Z. Haque}, \bibinfo{person}{L. Hong},
  \bibinfo{person}{V. Jain}, \bibinfo{person}{X. Liu}, {and}
  \bibinfo{person}{H. Shah}.} \bibinfo{year}{2016}\natexlab{}.
\newblock \showarticletitle{Wide \& Deep Learning for Recommender Systems}. In
  \bibinfo{booktitle}{\emph{Proc. 1st Workshop on Deep Learning for Recommender
  Systems}} (Boston, MA, USA) \emph{(\bibinfo{series}{DLRS 2016})}.
  \bibinfo{publisher}{ACM}, \bibinfo{address}{New York, NY, USA},
  \bibinfo{pages}{7–10}.
\newblock
\showISBNx{9781450347952}


\bibitem[Chowdhury et~al\mbox{.}(2017)]%
        {farseev3}
\bibfield{author}{\bibinfo{person}{Alok~Kumar Chowdhury},
  \bibinfo{person}{Aleksandr Farseev}, \bibinfo{person}{Prithwi~Raj
  Chakraborty}, \bibinfo{person}{Dian Tjondronegoro}, {and}
  \bibinfo{person}{Vinod Chandran}.} \bibinfo{year}{2017}\natexlab{}.
\newblock \showarticletitle{Automatic classification of physical exercises from
  wearable sensors using small dataset from non-laboratory settings}. In
  \bibinfo{booktitle}{\emph{2017 IEEE Life Sciences Conference (LSC)}}.
  \bibinfo{pages}{111--114}.
\newblock
\urldef\tempurl%
\url{https://doi.org/10.1109/LSC.2017.8268156}
\showDOI{\tempurl}


\bibitem[Devlin et~al\mbox{.}(2019)]%
        {devlin-etal-2019-bert}
\bibfield{author}{\bibinfo{person}{Jacob Devlin}, \bibinfo{person}{Ming-Wei
  Chang}, \bibinfo{person}{Kenton Lee}, {and} \bibinfo{person}{Kristina
  Toutanova}.} \bibinfo{year}{2019}\natexlab{}.
\newblock \showarticletitle{{BERT}: Pre-training of Deep Bidirectional
  Transformers for Language Understanding}. In \bibinfo{booktitle}{\emph{Proc.
  2019 NAACL}}. \bibinfo{publisher}{ACL}, \bibinfo{pages}{4171--4186}.
\newblock


\bibitem[Dosovitskiy et~al\mbox{.}(2020)]%
        {VIT}
\bibfield{author}{\bibinfo{person}{Alexey Dosovitskiy}, \bibinfo{person}{Lucas
  Beyer}, \bibinfo{person}{Alexander Kolesnikov}, \bibinfo{person}{Dirk
  Weissenborn}, \bibinfo{person}{Xiaohua Zhai}, \bibinfo{person}{Thomas
  Unterthiner}, \bibinfo{person}{Mostafa Dehghani}, \bibinfo{person}{Matthias
  Minderer}, \bibinfo{person}{Georg Heigold}, \bibinfo{person}{Sylvain Gelly},
  {et~al\mbox{.}}} \bibinfo{year}{2020}\natexlab{}.
\newblock \showarticletitle{An image is worth 16x16 words: Transformers for
  image recognition at scale}.
\newblock \bibinfo{journal}{\emph{arXiv preprint arXiv:2010.11929}}
  (\bibinfo{year}{2020}).
\newblock


\bibitem[Du et~al\mbox{.}(2023)]%
        {nie1}
\bibfield{author}{\bibinfo{person}{Yali Du}, \bibinfo{person}{Yinwei Wei},
  \bibinfo{person}{Wei Ji}, \bibinfo{person}{Fan Liu}, \bibinfo{person}{Xin
  Luo}, {and} \bibinfo{person}{Liqiang Nie}.} \bibinfo{year}{2023}\natexlab{}.
\newblock \showarticletitle{Multi-Queue Momentum Contrast for
  Microvideo-Product Retrieval}. In \bibinfo{booktitle}{\emph{Proceedings of
  the Sixteenth ACM International Conference on Web Search and Data Mining}}
  (Singapore, Singapore) \emph{(\bibinfo{series}{WSDM '23})}.
  \bibinfo{publisher}{ACM}, \bibinfo{address}{New York, NY, USA},
  \bibinfo{pages}{1003–1011}.
\newblock
\showISBNx{9781450394079}
\urldef\tempurl%
\url{https://doi.org/10.1145/3539597.3570405}
\showDOI{\tempurl}


\bibitem[Farseev(2023)]%
        {somin_responsibility}
\bibfield{author}{\bibinfo{person}{Aleksandr Farseev}.}
  \bibinfo{year}{2023}\natexlab{}.
\newblock \showarticletitle{Under the Hood of Social Media Advertising: How Do
  We Use AI Responsibly for Advertising Targeting and Creative Evaluation}. In
  \bibinfo{booktitle}{\emph{Proceedings of the Sixteenth ACM International
  Conference on Web Search and Data Mining}} (Singapore, Singapore)
  \emph{(\bibinfo{series}{WSDM '23})}. \bibinfo{publisher}{ACM},
  \bibinfo{address}{New York, NY, USA}, \bibinfo{pages}{1281–1282}.
\newblock
\showISBNx{9781450394079}
\urldef\tempurl%
\url{https://doi.org/10.1145/3539597.3575791}
\showDOI{\tempurl}


\bibitem[Farseev et~al\mbox{.}(2014)]%
        {farseev1}
\bibfield{author}{\bibinfo{person}{A Farseev}, \bibinfo{person}{N Gukov},
  \bibinfo{person}{I Gossoudarev}, {and} \bibinfo{person}{U Zarichnyak}.}
  \bibinfo{year}{2014}\natexlab{}.
\newblock \showarticletitle{Cross-platform online venue and user community
  recommendation based upon social networks data mining}.
\newblock \bibinfo{journal}{\emph{Computer Instruments in Education}}
  \bibinfo{volume}{6} (\bibinfo{year}{2014}), \bibinfo{pages}{28--38}.
\newblock


\bibitem[Farseev et~al\mbox{.}(2018)]%
        {somin_influencer}
\bibfield{author}{\bibinfo{person}{Aleksandr Farseev}, \bibinfo{person}{Kirill
  Lepikhin}, \bibinfo{person}{Hendrik Schwartz}, \bibinfo{person}{Eu~Khoon
  Ang}, {and} \bibinfo{person}{Kenny Powar}.} \bibinfo{year}{2018}\natexlab{}.
\newblock \showarticletitle{SoMin.Ai: Social Multimedia Influencer Discovery
  Marketplace}. In \bibinfo{booktitle}{\emph{Proceedings of the 26th ACM
  International Conference on Multimedia}} (Seoul, Republic of Korea)
  \emph{(\bibinfo{series}{MM '18})}. \bibinfo{publisher}{ACM},
  \bibinfo{address}{New York, NY, USA}, \bibinfo{pages}{1234–1236}.
\newblock
\showISBNx{9781450356657}
\urldef\tempurl%
\url{https://doi.org/10.1145/3240508.3241387}
\showDOI{\tempurl}


\bibitem[Farseev et~al\mbox{.}(2021)]%
        {we1}
\bibfield{author}{\bibinfo{person}{Aleksandr Farseev}, \bibinfo{person}{Qi
  Yang}, \bibinfo{person}{Andrey Filchenkov}, \bibinfo{person}{Kirill
  Lepikhin}, \bibinfo{person}{Yu-Yi Chu-Farseeva}, {and}
  \bibinfo{person}{Daron-Benjamin Loo}.} \bibinfo{year}{2021}\natexlab{}.
\newblock \showarticletitle{SoMin.Ai: Personality-Driven Content Generation
  Platform}. In \bibinfo{booktitle}{\emph{Proceedings of the 14th ACM
  International Conference on Web Search and Data Mining}}
  \emph{(\bibinfo{series}{WSDM '21})}. \bibinfo{publisher}{ACM},
  \bibinfo{address}{New York, NY, USA}, \bibinfo{pages}{890–893}.
\newblock
\showISBNx{9781450382977}
\urldef\tempurl%
\url{https://doi.org/10.1145/3437963.3441714}
\showDOI{\tempurl}


\bibitem[Fukui et~al\mbox{.}(2019)]%
        {fukui2018cvpr}
\bibfield{author}{\bibinfo{person}{H. Fukui}, \bibinfo{person}{T. Hirakawa},
  \bibinfo{person}{T. Yamashita}, {and} \bibinfo{person}{H. Fujiyoshi}.}
  \bibinfo{year}{2019}\natexlab{}.
\newblock \showarticletitle{Attention Branch Network: Learning of Attention
  Mechanism for Visual Explanation}.
\newblock \bibinfo{journal}{\emph{Computer Vision and Pattern Recognition}}
  (\bibinfo{year}{2019}), \bibinfo{pages}{10705--10714}.
\newblock


\bibitem[Ge et~al\mbox{.}(2018)]%
        {baba-image-matters}
\bibfield{author}{\bibinfo{person}{T. Ge}, \bibinfo{person}{H. Liu},
  \bibinfo{person}{P. Yi}, \bibinfo{person}{S. Huang}, \bibinfo{person}{Z.
  Zhang}, \bibinfo{person}{X. Zhu}, \bibinfo{person}{Y. Zhang},
  \bibinfo{person}{K. Gai}, \bibinfo{person}{L. Zhao}, \bibinfo{person}{G.
  Zhou}, \bibinfo{person}{K. Chen}, \bibinfo{person}{S. Liu},
  \bibinfo{person}{H. Yi}, \bibinfo{person}{Z. Hu}, \bibinfo{person}{B. Liu},
  {and} \bibinfo{person}{P. Sun}.} \bibinfo{year}{2018}\natexlab{}.
\newblock \showarticletitle{Image Matters: Visually Modeling User Behaviors
  Using Advanced Model Server}. \bibinfo{pages}{2087--2095}.
\newblock


\bibitem[He et~al\mbox{.}(2017)]%
        {nie3}
\bibfield{author}{\bibinfo{person}{Xiangnan He}, \bibinfo{person}{Lizi Liao},
  \bibinfo{person}{Hanwang Zhang}, \bibinfo{person}{Liqiang Nie},
  \bibinfo{person}{Xia Hu}, {and} \bibinfo{person}{Tat-Seng Chua}.}
  \bibinfo{year}{2017}\natexlab{}.
\newblock \showarticletitle{Neural Collaborative Filtering}. In
  \bibinfo{booktitle}{\emph{Proceedings of the 26th International Conference on
  World Wide Web}} (Perth, Australia) \emph{(\bibinfo{series}{WWW '17})}.
  \bibinfo{publisher}{International World Wide Web Conferences Steering
  Committee}, \bibinfo{address}{Republic and Canton of Geneva, CHE},
  \bibinfo{pages}{173–182}.
\newblock
\showISBNx{9781450349130}
\urldef\tempurl%
\url{https://doi.org/10.1145/3038912.3052569}
\showDOI{\tempurl}


\bibitem[He et~al\mbox{.}(2014)]%
        {fb-ctr-practical-lessons}
\bibfield{author}{\bibinfo{person}{X. He}, \bibinfo{person}{J. Pan},
  \bibinfo{person}{O. Jin}, \bibinfo{person}{T. Xu}, \bibinfo{person}{B. Liu},
  \bibinfo{person}{T. Xu}, \bibinfo{person}{Y. Shi}, \bibinfo{person}{A.
  Atallah}, \bibinfo{person}{R. Herbrich}, \bibinfo{person}{S. Bowers}, {and}
  \bibinfo{person}{J.~Q. Candela}.} \bibinfo{year}{2014}\natexlab{}.
\newblock \showarticletitle{Practical Lessons from Predicting Clicks on Ads at
  Facebook}. In \bibinfo{booktitle}{\emph{Proc. 8th International Workshop on
  Data Mining for Online Advertising}} \emph{(\bibinfo{series}{ADKDD'14})}.
  \bibinfo{publisher}{ACM}, \bibinfo{pages}{1–9}.
\newblock
\showISBNx{9781450329996}


\bibitem[Huang et~al\mbox{.}(2023)]%
        {we2}
\bibfield{author}{\bibinfo{person}{Alfred Huang}, \bibinfo{person}{Qi Yang},
  \bibinfo{person}{Sergey Nikolenko}, \bibinfo{person}{Marlo Ongpin},
  \bibinfo{person}{Ilia Gossoudarev}, \bibinfo{person}{Ngoc Yen~Duong},
  \bibinfo{person}{Kirill Lepikhin}, \bibinfo{person}{Sergey Vishnyakov},
  \bibinfo{person}{Yuyi Chu-Farseeva}, {and} \bibinfo{person}{Aleksandr
  Farseev}.} \bibinfo{year}{2023}\natexlab{}.
\newblock \showarticletitle{SoCraft: Advertiser-Level Predictive Scoring for
  Creative Performance on Meta}. In \bibinfo{booktitle}{\emph{Proceedings of
  the Sixteenth ACM International Conference on Web Search and Data Mining}}
  (Singapore, Singapore) \emph{(\bibinfo{series}{WSDM '23})}.
  \bibinfo{publisher}{ACM}, \bibinfo{address}{New York, NY, USA},
  \bibinfo{pages}{1132–1135}.
\newblock
\showISBNx{9781450394079}
\urldef\tempurl%
\url{https://doi.org/10.1145/3539597.3573032}
\showDOI{\tempurl}


\bibitem[Huang et~al\mbox{.}(2020)]%
        {huang2020tabtransformer}
\bibfield{author}{\bibinfo{person}{X. Huang}, \bibinfo{person}{A. Khetan},
  \bibinfo{person}{M. Cvitkovic}, {and} \bibinfo{person}{Z. Karnin}.}
  \bibinfo{year}{2020}\natexlab{}.
\newblock \showarticletitle{TabTransformer: Tabular Data Modeling Using
  Contextual Embeddings}.
\newblock  (\bibinfo{year}{2020}).
\newblock
\showeprint[arxiv]{2012.06678}~[cs.LG]


\bibitem[Koltcov et~al\mbox{.}(2014)]%
        {10.1145/2615569.2615680}
\bibfield{author}{\bibinfo{person}{Sergei Koltcov}, \bibinfo{person}{Olessia
  Koltsova}, {and} \bibinfo{person}{Sergey Nikolenko}.}
  \bibinfo{year}{2014}\natexlab{}.
\newblock \showarticletitle{Latent Dirichlet Allocation: Stability and
  Applications to Studies of User-Generated Content}. In
  \bibinfo{booktitle}{\emph{Proceedings of the 2014 ACM Conference on Web
  Science}} (Bloomington, Indiana, USA) \emph{(\bibinfo{series}{WebSci '14})}.
  \bibinfo{publisher}{ACM}, \bibinfo{address}{New York, NY, USA},
  \bibinfo{pages}{161–165}.
\newblock
\showISBNx{9781450326223}
\urldef\tempurl%
\url{https://doi.org/10.1145/2615569.2615680}
\showDOI{\tempurl}


\bibitem[McMahan et~al\mbox{.}(2013)]%
        {goog-ctr-view-from-trenches}
\bibfield{author}{\bibinfo{person}{H.~B. McMahan}, \bibinfo{person}{G. Holt},
  \bibinfo{person}{D. Sculley}, \bibinfo{person}{M. Young}, \bibinfo{person}{D.
  Ebner}, \bibinfo{person}{J. Grady}, \bibinfo{person}{L. Nie},
  \bibinfo{person}{T. Phillips}, \bibinfo{person}{E. Davydov},
  \bibinfo{person}{D. Golovin}, \bibinfo{person}{S. Chikkerur},
  \bibinfo{person}{D. Liu}, \bibinfo{person}{M. Wattenberg},
  \bibinfo{person}{A.~M. Hrafnkelsson}, \bibinfo{person}{T. Boulos}, {and}
  \bibinfo{person}{J. Kubica}.} \bibinfo{year}{2013}\natexlab{}.
\newblock \showarticletitle{Ad Click Prediction: A View from the Trenches}. In
  \bibinfo{booktitle}{\emph{Proc. 19th ACM SIGKDD}} \emph{(\bibinfo{series}{KDD
  '13})}. \bibinfo{publisher}{ACM}, \bibinfo{pages}{1222–1230}.
\newblock
\showISBNx{9781450321747}


\bibitem[Meta(2023)]%
        {metaBlueprints}
\bibfield{author}{\bibinfo{person}{Meta}.} \bibinfo{year}{2023}\natexlab{}.
\newblock \bibinfo{title}{Meta Blueprint}.
\newblock
\newblock
\urldef\tempurl%
\url{https://www.facebookblueprint.com/student/catalog}
\showURL{%
\tempurl}
\newblock
\shownote{Accessed on June 06, 2023}.


\bibitem[Nikolenko(2015)]%
        {10.1007/978-3-319-27101-9_5}
\bibfield{author}{\bibinfo{person}{Sergey Nikolenko}.}
  \bibinfo{year}{2015}\natexlab{}.
\newblock \showarticletitle{SVD-LDA: Topic Modeling for Full-Text Recommender
  Systems}. In \bibinfo{booktitle}{\emph{Advances in Artificial Intelligence
  and Its Applications}}, \bibfield{editor}{\bibinfo{person}{Obdulia
  Pichardo~Lagunas}, \bibinfo{person}{Oscar Herrera~Alc{\'a}ntara}, {and}
  \bibinfo{person}{Gustavo Arroyo~Figueroa}} (Eds.).
  \bibinfo{publisher}{Springer International Publishing},
  \bibinfo{address}{Cham}, \bibinfo{pages}{67--79}.
\newblock
\showISBNx{978-3-319-27101-9}


\bibitem[OpenAI(2023)]%
        {openai2023gpt4}
\bibfield{author}{\bibinfo{person}{OpenAI}.} \bibinfo{year}{2023}\natexlab{}.
\newblock \bibinfo{title}{GPT-4 Technical Report}.
\newblock
\newblock
\showeprint[arxiv]{2303.08774}~[cs.CL]


\bibitem[Ouyang et~al\mbox{.}(2022)]%
        {ouyang2022training}
\bibfield{author}{\bibinfo{person}{Long Ouyang}, \bibinfo{person}{Jeff Wu},
  \bibinfo{person}{Xu Jiang}, \bibinfo{person}{Diogo Almeida},
  \bibinfo{person}{Carroll~L. Wainwright}, \bibinfo{person}{Pamela Mishkin},
  \bibinfo{person}{Chong Zhang}, \bibinfo{person}{Sandhini Agarwal},
  \bibinfo{person}{Katarina Slama}, \bibinfo{person}{Alex Ray},
  \bibinfo{person}{John Schulman}, \bibinfo{person}{Jacob Hilton},
  \bibinfo{person}{Fraser Kelton}, \bibinfo{person}{Luke Miller},
  \bibinfo{person}{Maddie Simens}, \bibinfo{person}{Amanda Askell},
  \bibinfo{person}{Peter Welinder}, \bibinfo{person}{Paul Christiano},
  \bibinfo{person}{Jan Leike}, {and} \bibinfo{person}{Ryan Lowe}.}
  \bibinfo{year}{2022}\natexlab{}.
\newblock \bibinfo{title}{Training language models to follow instructions with
  human feedback}.
\newblock
\newblock
\showeprint[arxiv]{2203.02155}~[cs.CL]


\bibitem[Ouyang et~al\mbox{.}(2019)]%
        {baba-rep-learning-ctr}
\bibfield{author}{\bibinfo{person}{Wentao Ouyang}, \bibinfo{person}{Xiuwu
  Zhang}, \bibinfo{person}{Shukui Ren}, \bibinfo{person}{Chao Qi},
  \bibinfo{person}{Zhaojie Liu}, {and} \bibinfo{person}{Yanlong Du}.}
  \bibinfo{year}{2019}\natexlab{}.
\newblock \showarticletitle{Representation Learning-Assisted Click-Through Rate
  Prediction}. In \bibinfo{booktitle}{\emph{Proc. 28th IJCAI}}.
  \bibinfo{pages}{4561--4567}.
\newblock
\urldef\tempurl%
\url{https://doi.org/10.24963/ijcai.2019/634}
\showDOI{\tempurl}


\bibitem[Ricci et~al\mbox{.}(2010)]%
        {10.5555/1941884}
\bibfield{author}{\bibinfo{person}{Francesco Ricci}, \bibinfo{person}{Lior
  Rokach}, \bibinfo{person}{Bracha Shapira}, {and} \bibinfo{person}{Paul~B.
  Kantor}.} \bibinfo{year}{2010}\natexlab{}.
\newblock \bibinfo{booktitle}{\emph{Recommender Systems Handbook}
  (\bibinfo{edition}{1st} ed.)}.
\newblock \bibinfo{publisher}{Springer-Verlag}, \bibinfo{address}{Berlin,
  Heidelberg}.
\newblock
\showISBNx{0387858199}


\bibitem[Richardson et~al\mbox{.}(2007)]%
        {msft-ad-pred}
\bibfield{author}{\bibinfo{person}{Matthew Richardson}, \bibinfo{person}{Ewa
  Dominowska}, {and} \bibinfo{person}{Robert Ragno}.}
  \bibinfo{year}{2007}\natexlab{}.
\newblock \showarticletitle{Predicting Clicks: Estimating the Click-through
  Rate for New Ads}. In \bibinfo{booktitle}{\emph{Proc. 16th WWW}}
  \emph{(\bibinfo{series}{WWW '07})}. \bibinfo{publisher}{ACM},
  \bibinfo{pages}{521–530}.
\newblock
\showISBNx{9781595936547}


\bibitem[Rombach et~al\mbox{.}(2022)]%
        {stablediffusion}
\bibfield{author}{\bibinfo{person}{Robin Rombach}, \bibinfo{person}{Andreas
  Blattmann}, \bibinfo{person}{Dominik Lorenz}, \bibinfo{person}{Patrick
  Esser}, {and} \bibinfo{person}{Bj\"orn Ommer}.}
  \bibinfo{year}{2022}\natexlab{}.
\newblock \showarticletitle{High-Resolution Image Synthesis With Latent
  Diffusion Models}. In \bibinfo{booktitle}{\emph{Proceedings of the IEEE/CVF
  Conference on Computer Vision and Pattern Recognition (CVPR)}}.
  \bibinfo{pages}{10684--10695}.
\newblock


\bibitem[Savchenko et~al\mbox{.}(2020)]%
        {savchenko-etal-2020-ad}
\bibfield{author}{\bibinfo{person}{Andrey Savchenko}, \bibinfo{person}{Anton
  Alekseev}, \bibinfo{person}{Sejeong Kwon}, \bibinfo{person}{Elena
  Tutubalina}, \bibinfo{person}{Evgeny Myasnikov}, {and}
  \bibinfo{person}{Sergey Nikolenko}.} \bibinfo{year}{2020}\natexlab{}.
\newblock \showarticletitle{Ad Lingua: Text Classification Improves Symbolism
  Prediction in Image Advertisements}. In \bibinfo{booktitle}{\emph{Proceedings
  of the 28th International Conference on Computational Linguistics}}.
  \bibinfo{publisher}{International Committee on Computational Linguistics},
  \bibinfo{address}{Barcelona, Spain (Online)}, \bibinfo{pages}{1886--1892}.
\newblock
\urldef\tempurl%
\url{https://doi.org/10.18653/v1/2020.coling-main.171}
\showDOI{\tempurl}


\bibitem[Shenbin et~al\mbox{.}(2020)]%
        {10.1145/3336191.3371831}
\bibfield{author}{\bibinfo{person}{Ilya Shenbin}, \bibinfo{person}{Anton
  Alekseev}, \bibinfo{person}{Elena Tutubalina}, \bibinfo{person}{Valentin
  Malykh}, {and} \bibinfo{person}{Sergey~I. Nikolenko}.}
  \bibinfo{year}{2020}\natexlab{}.
\newblock \showarticletitle{RecVAE: A New Variational Autoencoder for Top-N
  Recommendations with Implicit Feedback}. In
  \bibinfo{booktitle}{\emph{Proceedings of the 13th International Conference on
  Web Search and Data Mining}} (Houston, TX, USA) \emph{(\bibinfo{series}{WSDM
  '20})}. \bibinfo{publisher}{ACM}, \bibinfo{address}{New York, NY, USA},
  \bibinfo{pages}{528–536}.
\newblock
\showISBNx{9781450368223}
\urldef\tempurl%
\url{https://doi.org/10.1145/3336191.3371831}
\showDOI{\tempurl}


\bibitem[Tutubalina and Nikolenko(2017)]%
        {TN17}
\bibfield{author}{\bibinfo{person}{Elena Tutubalina} {and}
  \bibinfo{person}{Sergey~I. Nikolenko}.} \bibinfo{year}{2017}\natexlab{}.
\newblock \showarticletitle{Demographic Prediction based on User Reviews about
  Medications}.
\newblock \bibinfo{journal}{\emph{Computación y Sistemas}}
  \bibinfo{volume}{21}, \bibinfo{number}{2} (\bibinfo{year}{2017}),
  \bibinfo{pages}{227--241}.
\newblock


\bibitem[Tutubalina and Nikolenko(2018)]%
        {TN18}
\bibfield{author}{\bibinfo{person}{Elena Tutubalina} {and}
  \bibinfo{person}{Sergey~I. Nikolenko}.} \bibinfo{year}{2018}\natexlab{}.
\newblock \showarticletitle{Exploring convolutional neural networks and topic
  models for user profiling from drug reviews}.
\newblock \bibinfo{journal}{\emph{Multimedia Tools and Applications}}
  \bibinfo{volume}{77}, \bibinfo{number}{4} (\bibinfo{year}{2018}),
  \bibinfo{pages}{4791--4809}.
\newblock


\bibitem[Wang et~al\mbox{.}(2022)]%
        {nie2}
\bibfield{author}{\bibinfo{person}{Wenjie Wang}, \bibinfo{person}{Fuli Feng},
  \bibinfo{person}{Liqiang Nie}, {and} \bibinfo{person}{Tat-Seng Chua}.}
  \bibinfo{year}{2022}\natexlab{}.
\newblock \showarticletitle{User-Controllable Recommendation Against Filter
  Bubbles}. In \bibinfo{booktitle}{\emph{Proceedings of the 45th International
  ACM SIGIR Conference on Research and Development in Information Retrieval}}
  (Madrid, Spain) \emph{(\bibinfo{series}{SIGIR '22})}.
  \bibinfo{publisher}{ACM}, \bibinfo{address}{New York, NY, USA},
  \bibinfo{pages}{1251–1261}.
\newblock
\showISBNx{9781450387323}
\urldef\tempurl%
\url{https://doi.org/10.1145/3477495.3532075}
\showDOI{\tempurl}


\bibitem[Yang et~al\mbox{.}(2021)]%
        {rec4}
\bibfield{author}{\bibinfo{person}{Qi Yang}, \bibinfo{person}{Aleksandr
  Farseev}, {and} \bibinfo{person}{Andrey Filchenkov}.}
  \bibinfo{year}{2021}\natexlab{}.
\newblock \showarticletitle{Two-Faced Humans on Twitter and Facebook:
  Harvesting Social Multimedia for Human Personality Profiling}. In
  \bibinfo{booktitle}{\emph{Proceedings of the 2021 Workshop on Intelligent
  Cross-Data Analysis and Retrieval}} (Taipei, Taiwan)
  \emph{(\bibinfo{series}{ICDAR '21})}. \bibinfo{publisher}{ACM},
  \bibinfo{address}{New York, NY, USA}, \bibinfo{pages}{39–47}.
\newblock
\showISBNx{9781450385299}
\urldef\tempurl%
\url{https://doi.org/10.1145/3463944.3469270}
\showDOI{\tempurl}


\bibitem[Yang et~al\mbox{.}(2022a)]%
        {rec3}
\bibfield{author}{\bibinfo{person}{Qi Yang}, \bibinfo{person}{Aleksandr
  Farseev}, \bibinfo{person}{Sergey Nikolenko}, {and} \bibinfo{person}{Andrey
  Filchenkov}.} \bibinfo{year}{2022}\natexlab{a}.
\newblock \showarticletitle{Do we behave differently on Twitter and Facebook:
  Multi-view social network user personality profiling for content
  recommendation}.
\newblock \bibinfo{journal}{\emph{Frontiers in Big Data}}  \bibinfo{volume}{5}
  (\bibinfo{year}{2022}).
\newblock
\showISSN{2624-909X}
\urldef\tempurl%
\url{https://doi.org/10.3389/fdata.2022.931206}
\showDOI{\tempurl}


\bibitem[Yang et~al\mbox{.}(2022b)]%
        {10.1145/3503161.3548769}
\bibfield{author}{\bibinfo{person}{Qi Yang}, \bibinfo{person}{Sergey
  Nikolenko}, \bibinfo{person}{Alfred Huang}, {and} \bibinfo{person}{Aleksandr
  Farseev}.} \bibinfo{year}{2022}\natexlab{b}.
\newblock \showarticletitle{Personality-Driven Social Multimedia Content
  Recommendation}. In \bibinfo{booktitle}{\emph{Proceedings of the 30th ACM
  International Conference on Multimedia}} (Lisboa, Portugal)
  \emph{(\bibinfo{series}{MM '22})}. \bibinfo{publisher}{ACM},
  \bibinfo{address}{New York, NY, USA}, \bibinfo{pages}{7290–7299}.
\newblock
\showISBNx{9781450392037}
\urldef\tempurl%
\url{https://doi.org/10.1145/3503161.3548769}
\showDOI{\tempurl}


\bibitem[Yang et~al\mbox{.}(2023)]%
        {10.1145/3539597.3573031}
\bibfield{author}{\bibinfo{person}{Qi Yang}, \bibinfo{person}{Christos
  Tzelepis}, \bibinfo{person}{Sergey Nikolenko}, \bibinfo{person}{Ioannis
  Patras}, {and} \bibinfo{person}{Aleksandr Farseev}.}
  \bibinfo{year}{2023}\natexlab{}.
\newblock \showarticletitle{"Just To See You Smile": SMILEY, a Voice-Guided
  <Strike>GUY</Strike> GAN}. In \bibinfo{booktitle}{\emph{Proceedings of the
  Sixteenth ACM International Conference on Web Search and Data Mining}}
  (Singapore, Singapore) \emph{(\bibinfo{series}{WSDM '23})}.
  \bibinfo{publisher}{ACM}, \bibinfo{address}{New York, NY, USA},
  \bibinfo{pages}{1196–1199}.
\newblock
\showISBNx{9781450394079}
\urldef\tempurl%
\url{https://doi.org/10.1145/3539597.3573031}
\showDOI{\tempurl}


\end{thebibliography}

\end{document}